\documentclass[12pt]{iopart}

\usepackage{graphicx}

\usepackage{subfig}
\usepackage{epstopdf}

\begin{document}

\title[Local Unitary Equivalent Classes of Symmetric $N$-Qubit Mixed States]{Local Unitary Equivalent Classes of Symmetric $N$-Qubit Mixed States}
\author{Sakineh Ashourisheikhi \,and \, Swarnamala Sirsi}
\address{Department of Physics, Yuvaraja's College \\
University of Mysore, Mysore-570005,\,India}
\ead{Samin$_{-}$ashuri@yahoo.com}
\begin{abstract}
Majorana Representation (MR) of symmetric $ N $-qubit pure states has been used successfully in entanglement classification. Generalization of this has been a long standing open problem due to the difficulties faced in the construction of a Majorana like geometric representation for symmetric mixed state. We have overcome this problem by developing a method of classifying local unitary (LU) equivalent classes of symmetric $N$-qubit mixed states based on the geometrical Multiaxial Representation (MAR) of the density matrix. In addition to the two parameters defined for the entanglement classification of the symmetric pure states based on MR, namely, diversity degree and degeneracy configuration, we show that another parameter called rank needs to be introduced for symmetric mixed state classification. Our scheme of classification is more general as it can be applied to both pure and mixed states. To bring out the similarities/ differences between the MR and MAR, $ N $-qubit GHZ state is taken up for a detailed study. We conclude that pure state classification based on MR is not a special case of our classification scheme based on MAR. We also give a recipe to identify the most general symmetric $N$-qubit pure separable states. The power of our method is demonstrated using several well known examples of symmetric two qubit pure and mixed states as well as three qubit pure states. Classification of uniaxial, Biaxial and triaxial symmetric two qubit mixed states which can be produced in the laboratory is studied in detail. 

\end{abstract}

\maketitle

\section{Introduction}

Two density matrices $ \rho $ and $ \rho^{\prime} $ are said to be LU equivalent if ${\rho^{\prime}}$ = $U\rho U^{\dagger}$ where $U \in SU(2)^{\times{N} }$. LU equivalence of multipartite pure states has received a lot of attention recently in the context of the study of entangled states $^{1,2,3,4}  $. It is well known that the states belonging to the same LU equivalent class can be used for similar kind of quantum information processing tasks as they posses the same amount of entanglement. Different LU equivalent classes of up to pure 5-qubit states and for a few mixed states have been determined by introducing a standard form for multipartite states $^{2,3}$.
 
  One way of studying unitary equivalence of multipartite mixed states is, using the singular value decomposition method $ ^{5} $ and another way is by evaluating the LU invariants (LUI). There exist well known algebraic methods for generation of invariants  $ ^{6,7,8,9} $. As the number of subsystems of multipartite state increases, the problem of identifying and interpreting the independent invariants rapidly becomes very complicated. However, LUI associated with the symmetric states, which are experimentally viable and mathematically elegant, are easier to handle as the dimensionality  of the Hilbert space involved is much less. Because of the permutational symmetry involved in the symmetric state, $ \rho $ and ${\rho^{\prime}}$ are said to belong to the same LU equivalence class if 
\begin{equation}
\rho^{\prime}=R\otimes R\otimes R...R\,\rho\,R^{-1}\otimes R^{-1}...R^{-1},
\end{equation}
where $ R $ represents the rotation operator on a qubit $ ^{10,11} $. Sirsi and Adiga $ ^{12} $ have constructed LUI of the most general symmetric $ N $-qubit mixed systems using the elegant MAR$ ^{13} $ of the symmetric states. Entanglement classification of symmetric $ N $-qubit pure state has been studied by Bastin et. al$ ^{14} $ based on the well known MR$ ^{15} $. Generalization of this classification scheme to cover the mixed states is a difficult problem as one needs a Majorana like geometrical representation for a $ N $-qubit mixed state. Such a geometrical realization of a symmetric mixed state is deemed to be difficult as the number of parameters required to characterize $ N $-qubit mixed state namely $ N(N+2) $ is much larger than the $ 2N $ parameters required to characterize $ N $-qubit pure state$ ^{16,17} $. This problem was solved by Ramachandran and Ravishankar$ ^{13} $ by constructing a MAR of the $ 2j $ Fano statistical tensor parameters which characterize the spin-$ j $ assembly.
 Entanglement classification of mixed state, symmetric or otherwise, under LU transformation poses a difficult problem as the definition of mixed state entanglement itself is poorly understood. In this paper we propose a scheme for classifying the most general symmetric $N$-qubit mixed states under LU transformation by evaluating the LUI based on the MAR of Ramachandran and Ravishankar $ ^{13} $.

  The paper is organized as follows: In Sec. 2 we give a brief introduction to the symmetric states. MR of pure states and classification of pure states by Bastin et. al. $ ^{14} $ based on MR is given in sections 2.1 and 2.2 respectively. In sec.3, we introduce a standard expression for spin-j density matrix in terms of Fano statistical tensor parameters. A brief discussion of the method of obtaining the axes characterizing the density matrix using MAR is given in section 3.1. In sec. 3.2 we propose a method of LU equivalent classification of density matrices based on the MAR. The similarities/ differences between the MR and MAR are studied in detail for $  N$-qubit GHZ state in sec 4. A recipe to identify the most general $ N $-qubit separable state is given in sec. 5.1. The power of our method is demonstrated using several well known examples of symmetric
two qubit pure and mixed, states as well as three qubit
pure states in Sec 5.2 and 5.3. 

\section{ Symmetric State}
$ N $-qubits which obey permutational symmetry are represented by symmetric states which reside in the $ (N+1) $ dimensional subspace of the $ 2^{N} $ dimensional state space of $ N $-qubits. These states are experimentally viable$ ^{18,19,20} $ and offer elegant mathematical analysis as they are identified with the angular momentum states. An $ N $-qubit symmetric space can be considered to be spanned by the eigen states $  | jm\rangle $; $ m=-j,...,+j $ of angular momentum operators $ J^{2} $ and $ J_{z} $ with $ j=\frac{N}{2} $.

  Majorana's geometric realization$ ^{15} $ of a spin-$ j $ state as a constellation of  $ 2j=N $ points on Block sphere has been used extensively in recent times$ ^{21,22} $. But the equally elegant MAR of the most general spin-$ j $ state$ ^{13} $, both pure and mixed, is not very well known. For the sake of completeness we describe  both the representations briefly in the subsequent sections.
\subsection{Majorana Representation of Pure States}
   We know that the most general spin-$ j$ pure state $ | \psi^{j}\rangle $ is given by
\begin{equation}
| \psi^{j}\rangle=\sum_{m=-j}^{+j} a_{m}\,| jm\rangle.
\end{equation} 
Let us consider a rotation $  R(\phi,\theta,0)$ of the frame of reference such that the expansion coefficient $ a_{-j} $ in the rotated frame vanishes i.e;
\[(a_{-j})^{R}=0=\langle{j-j}|R^{-1}(\phi,\theta,0)| \psi_{j}\rangle=\sum_{m} a_{m}\langle{j-j}|R^{-1}(\phi,\theta,0)| jm\rangle\]
\begin{equation}
=\sum_{m} a_{m}\, D^{\ast_{j}}_{m\,-j}(\phi,\theta,0)=\sum_{m} a_{m}\,(-1)^{-(j+m)}\, D^{{j}}_{-m\,j}(\phi,\theta,0),
\end{equation}
where $ D^{j}_{m^{\prime}\,m}(\phi,\theta,0) $ are the matrix elements of Wigner rotation matrices, given by$^{23} $

\[ D^{j}_{m^{\prime}\, m}(\alpha\beta\gamma)=e^{-im^{\prime}\alpha}e^{-im\gamma}\sum_{s}\frac{(-1)^{s}\sqrt{(j+m)!(j-m)!(j+m^{\prime})!(j-m^{\prime})!}}{s!(j-s-m^{\prime})!(j+m-s)!(m^{\prime}+s-m)!}\]
 \begin{equation}
 \times \left(   cos\frac{\beta}{2}\right)^{2j+m-m^{\prime}-2s}\,\,(-1)^{m^{\prime}-m+2s}\, \left( sin\frac{\beta}{2}\right)^{m^{\prime}-m+2s}.   
\end{equation}
Thus
 \begin{equation}
D^{j}_{-m\,j}(\phi,\theta,0)=e^{im\phi}\,(-1)^{j+m}\sqrt{^{2j}C_{j+m}}\left(   cos\frac{\theta}{2}\right)^{j-m}\,\,(-1)^{j+m}\, \left( sin\frac{\theta}{2}\right)^{j+m}
\end{equation}
where $ s=j+m $ and $ ^{2j}C_{j+m} $ is the Binomial Coefficient. Eq. (3) becomes
\begin{equation}
\mathcal{A}\sum_{m=-j}^{+j}(-1)^{j+m}\sqrt{^{2j}\,C_{j+m}}\,a_{m}\,Z^{j+m}=0
\end{equation}
where $ Z=tan\left( \frac{\theta}{2}\right)\,e^{i\phi}  $ and the overall coefficient $  \mathcal{A}=cos^{2j}\left( \frac{\theta}{2}\right)\,e^{-i\phi j}$.\\
The Majorana polynomial $ P(z) $ is given by
\begin{equation}
P(Z)=\sum_{m=-j}^{+j}(-1)^{j+m}\sqrt{^{2j}\,C_{j+m}}\,a_{m}\,Z^{j+m}=0,
\end{equation}
for $ \theta\neq \pi $.\\
Equivalently, from eq. (5) we can also have 
\begin{equation}
\mathcal{A^{\prime}}\sum_{m=-j}^{+j}(-1)^{j+m}\sqrt{^{2j}\,C_{j+m}}\,a_{m}\, Z^{\prime ^{j-m}} =0
\end{equation}
where $  \mathcal{A^{\prime}}=sin^{2j}\left( \frac{\theta}{2}\right)\,e^{i\phi j}$ and $ Z^{\prime}=\frac{1}{Z}=cot\left( \frac{\theta}{2}\right) \,e^{-i\phi} $.

 We thus obtain,
 \begin{equation}
P(Z^{\prime})=\sum_{m=-j}^{+j}(-1)^{j-m}\sqrt{^{2j}\,C_{j+m}}\,a_{m}\,Z^{\prime^{j-m}}=0
\end{equation} 
 for $ \theta\neq 0 $.\\
  Solving either of the polynomial equations, one gets $ 2j $ solutions namely $\{(\theta_{1},\phi_{1}),(\theta_{2},\phi_{2}),...,(\theta_{2j},\phi_{2j})\} $ in general.
   Thus every pure spin-$ j $ state $ | \psi^{j}\rangle $ or the corresponding symmetric state $ | \psi^{2j}_{sym}\rangle $ can be represented by a constellation of $ 2j $ points on the Block sphere or   
\begin{equation}
| \psi^{2j}_{sym}\rangle= \mathcal{N}\sum_{P} \hat{P}(\epsilon_{1},\epsilon_{2},...,\epsilon_{2j}),
\end{equation}  
where 
\begin{equation}
| \epsilon_{k}\rangle=cos(\theta_{k}/2)e^{-i\phi_{k}/2}\,| 0\rangle+sin(\theta_{k}/2)e^{i\phi_{k}/2}\,| 1\rangle,\,\,\,\, k=0,1,...,2j
\end{equation}
refer to the $N $ spinors constituting the symmetric state $ | \psi^{2j}_{sym}\rangle $; $ \hat{P} $ corresponds to the set of $ (2j)! $ permutations of the spinors and $ \mathcal{N} $ corresponds to an overall normalization factor. 
 \subsection{ Classification of Pure States }
Entanglement classification of qubits based on  SL(2,C) $^{24} $, LU $ ^{2,3} $, Stochastic Local Operatoion and Classical Communication   (SLOCC) $^{16,17,25,26}$ and Local Operation and Classical Communication (LOCC) $^{27}$ has gained importance in recent times. The SLOCC classification of the permutationally symmetric $ N $-qubit states makes use of the elegant, geometrical representation of the spin-$ j $  states given by Majorana. 
According to SLOCC classification of symmetric pure states by Bastin et. al.$ ^{14} $, the number of identical spinors $ | \epsilon_{i}\rangle  $ in eq. (10) is called the $ degeneracy$ $number $. Further, the $ degeneracy $ $configuration $ $\mathcal{D}_{\{n_{i}\}}$ of a symmetric state $ | \psi^{2j}_{sym}\rangle $ is defined such that $ {\{n_{i}\}} $ is the set of degeneracy numbers ordered in decreasing order by convention. The number of $ n_{i}\,{'}s $ defines the diversity degree of the symmetric state.
    For example, if all the $ N $ spinors of a symmetric $ N $-qubit pure state are identical, then the state is said to have the degeneracy configuration $  \mathcal{D}_{N} $ and diversity degree $ d=1 $. Similarly, if all except two spinors are identical then the state has the degeneracy configuration of $ \mathcal{D}_{N-2,2} $ and  $ d=2 $ or $ \mathcal{D}_{N-2,1,1}  , d=3 $ depending on whether the two remaining spinors are identical or not respectively. Thus a separable symmetric $ N $-qubit pure state has the degeneracy configuration of $\mathcal{D}_{N} $ and $ d=1 $. 
 
 Such a classification based on MR is valid for symmetric pure states only. Therefore we propose a novel scheme for the most general symmetric $ N $-qubit pure as well as mixed states based on an equally elegant MAR of the density matrix. 
  
 \section{Fano Representation of Spin-$ j $ Assembly}
 We begin by introducing the spherical tensor representation of the most general, $ N $-Qubit pure as well as mixed systems. A standard expression$ ^{28,29,30} $ for the most general spin-j density matrix in terms of Fano statistical tensor parameters $t^{k}_{q}\,{'}s$ 
is given by 
\begin{equation}
\rho(\vec{J}) = \frac {Tr(\rho)}{(2j+1)}\sum^{2j}_{k=0}\,\sum^{+k}_{q=-k}\,\, t^{k}_{q}\, \tau^{k^{\dagger}}_{q}(\vec{J})\,\,,
\end{equation} 
where $ \vec{J} $ is the angular momentum operator with components $ J_{x}, J_{y}, J_{z} $ and $\tau^{k}_{q}\,{'}s$ \, (with $\tau^{0}_{0} = I$ ,the identity operator) are irreducible tensor operators of rank $k$ in the $ 2j+1$ dimensional 
spin space with projection $q$ along the axis of quantization in the real 3-dimensional space. The matrix elements of $\tau^{k}_{q}$ are given by 
\begin{equation}
\langle{jm'}|\tau^{k}_{q}(\vec{J})|{jm}\rangle = [k]\,\,C(jkj;mqm')
\end{equation}
 where $C(jkj;mqm')$ are the Clebsch-Gordan coefficients and $ [k]=\sqrt{2k+1} $. The $\tau^{k}_{q}\,{'}s$ satisfy the orthogonality relations
\begin{equation}
Tr({\tau^{k^{\dagger}}_{q}\tau^{k^{'}}_{q^{'}}})= (2j+1)\,\delta_{kk^{'}} \delta_{qq^{'}}\,
\end{equation}
where $ \tau^{k^{\dagger}}_{q}=(-1)^{q}\,\tau^{k}_{q} $. Here the normalization has been chosen so as to be in agreement with Madison convention $ ^{31} $.\\
The irreducible tensor operators  ${\tau^{k}_{q}}\,{'}s$ have simple transformation properties under co-ordinate rotation in the 3-dimensional space. In the rotated frame  ${\tau^{k}_{q}}\,{'}s$ are given by\begin{equation}
(\tau^{k}_{q})^{R} = \sum^{+k}_{q^{'}=-k}\,\, D^{k}_{q^{'}q}(\phi,\theta,\psi)\,\tau^{k}_{q^{'}}\,\,,
\end{equation}
where $D^{k}_{q^{'}q}(\phi,\theta,\psi)$ denote Wigner-$D$ matrices parametrized by Euler angles $(\phi, \theta, \psi)$. The spherical tensor parameters $t^{k}_{q}\,{'}s$
which characterize the given system are the average expectation values
\begin{equation}
t^{k}_{q} =  Tr(\rho\,\tau^{k}_{q})=\sum_{m=-j}^{+j}\,\langle{jm}|\rho\,\tau^{k}_{q}|{jm}\rangle.
\end{equation} 
Using eq. (13), 
\begin{equation}
t^{k}_{q}=\sum_{m=-j}^{+j}\,\rho_{mm^{\prime}}\,[k]\,C(jkj;mqm^{\prime}),\,\,\,\,\,\,\,\,\, with\,\,\,\, m^{\prime}=m+q.
\end{equation}
Note that $ t^{0}_{0}=1 $. Since $\rho$ is Hermitian and $\tau^{k^{\dagger}}_{q} = (-1)^{q}\tau^{k}_{-q}$,  complex conjugate of $t^{k}_{q}\,{'}s$ satisfy the condition 
 \begin{equation}
t^{k^{*}}_{q}=(-1)^{q}\,t^{k}_{-q}\,\,.
 \end{equation}
Also, $\rho=\rho^{\dagger}$ and $Tr\rho=1$ imply that $\rho$ can be specified by    $n^{2}-1$ independent parameters where $n=2j+1$ is the dimension of the Hilbert space.\\
In the rotated frame ${t^{k}_{q}}\,{'}s$ are given by 
\begin{equation}
(t^{k}_{q})^{R} = \sum^{+k}_{q^{'}=-k}\,\, D^{k}_{q^{'}q}(\phi,\theta,\psi)\,t^{k}_{q^{'}}\,\,.
\end{equation} 
 \subsection{ Multiaxial Representation of Spin-$ j $ Systems$ ^{13} $}
In the case of most general spin-$ j $ state, both pure as well as mixed, $ \rho(\vec{J}) $ is given by eq. (12). Thus $ \rho $ can be parametrized in terms of $ 2j+1 $ spherical tensors $ t^{k}_{q} $; $ k=0,1,2,...2j $ , $ q=-k $ to $ +k $. Consider a rotation $ R(\phi,\theta,0)$ of the frame of reference such that $ t^{k}_{k} $ in the rotated frame vanishes i.e from eq. (19)
\begin{equation}
(t^{k}_{k})^{R} =0= \sum_{q=-k}^{+k}\,\, D^{k}_{qk}(\phi,\theta,0)\,t^{k}_{q}\,\,.
\end{equation}
Again using the Wigner expression for $ D^{j} $ matrices, eq. (4), in eq. (20), we obtain 
\begin{equation}
\sum_{q=-k}^{k}\,  e^{-iq\phi}\,(-1)^{k-q}\,\sqrt{^{2k}\,C_{k+q}}\,\,t^{k}_{q}\,\left(  cos\frac{\theta}{2}\right)^{k+q} \,(-1)^{k-q}\,\left(  sin\frac{\theta}{2}\right)^{k-q} =0
\end{equation}
or
\begin{equation}
\mathcal{A}\,\sum_{q=-k}^{+k}\,(-1)^{2(k-q)}\,\sqrt{^{2k}\,C_{k+q}}\,\,t^{k}_{q}\,Z^{k-q}=0,
\end{equation}
where $ Z=tan\left( \frac{\theta}{2}\right)\,e^{i\phi}  $ and the overall coefficient $  \mathcal{A}=cos^{2k}\left( \frac{\theta}{2}\right)\,e^{-ik\phi}$.\\
Thus the polynomial equation $ P(Z) $ is given by
\begin{equation}
P(Z)=\sum_{q=-k}^{+k}(-1)^{2(k-q)}\sqrt{^{2k}\,C_{k+q}}\,t^{k}_{q}\,Z^{k-q}=0,
\end{equation} 
 which for every $ k $ leads to $ 2k $ solutions namely $\{(\theta_{1},\phi_{1}),(\theta_{2},\phi_{2}),...,(\theta_{k},\phi_{k}),(\pi-\theta_{1},\pi+\phi_{1}),...,(\pi-\theta_{k},\pi+\phi_{k})\}  $. Thus the $ 2k $ solutions constitute $ k $ axes or $ k $ double headed arrows. Therefore, every $ t^{k}_{q} $ can be constructed as follows: 
\begin{equation}
t^{k}_{q} = r_{k}(...((\hat{Q}(\theta_{1},\phi_{1})\otimes\hat{Q}(\theta_{2},\phi_{2}))^{2}\otimes\hat{Q}(\theta_{3},\phi_{3}))^{3}\otimes...)^{k-1}\otimes\hat{Q}(\theta_{k},\phi_{k}))^{k}_{q} \, 
\end{equation}
where 
\begin{equation}
 (\hat{Q}(\theta_{1},\phi_{1})\otimes\hat{Q}(\theta_{2},\phi_{2}))^{2}_{q}=\sum _{q_{1}}C(11k;q_{1}q_{2}q)(\hat{Q}(\theta_{1},\phi_{1}))^{1}_{q_{1}} (\hat{Q}(\theta_{2},\phi_{2}))^{1}_{q_{2}}
 \end{equation}
 and the spherical components of $ \hat{Q} $ are given by,
\begin{equation}
(\hat{Q}(\theta,\phi))^{1}_{q}= \sqrt{\frac{4\pi}{3}}\,\,Y^{1}_{q}(\theta,\phi).
\end{equation}
Here $Y^{1}_{q}(\theta,\phi)$ are the well known spherical harmonics.\\

  Observe that from eq. (21) we can also have 
\begin{equation}
\mathcal{A^{\prime}}\,\sum_{q=-k}^{+k}(-1)^{2(k-q)}\sqrt{^{2k}\,C_{k+q}}\,t^{k}_{q}\,Z^{\prime^{k+q}}=0
\end{equation}
where $  \mathcal{A^{\prime}}=sin^{2k}\left( \frac{\theta}{2}\right)\,e^{ik\phi}$ and $ Z^{\prime}=\frac{1}{Z}=cot\left( \frac{\theta}{2}\right) \,e^{-i\phi} $. Therefore,
\begin{equation}
P(Z^{\prime})=\sum_{q=-k}^{+k}(-1)^{2(k-q)}\sqrt{^{2k}\,C_{k+q}}\,t^{k}_{q}\,Z^{\prime^{k+q}}=0
\end{equation} 
for every $ k $, leading to the same set of $ 2k $ solutions as obtained from eq. (23).
 
  Thus in MAR, the symmetric state of $N$-qubit assembly can be represented geometrically by a set of $N=2j$ spheres of different radii $r_{1},r_{2},...,r_{k}$ corresponding to each value of $k$. The $k^{th}$ sphere in general consists of a constellation of 2$k$ points on its surface specified by $\hat{Q}(\theta_{i},\phi_{i})$ and   $\hat{Q}(\pi-\theta_{i},\pi+\phi_{i})$; $i=1,2,...,k $. In other words, every $ t^{k} $ is specified by $ k $ axes in a sphere of radius $ r_{k} $. \\

  On the other hand, consider a spin-$ j $ density matrix which is characterized by non-zero $ t^{k}_{0} $ $ (k=1,2...2j) $ only. In a rotated frame we have from eq. (19)
\begin{equation}
(t^{k}_{q})^{R}=D^{k}_{0q}(\phi,\theta,0)\,t^{k}_{0}.
\end{equation}
 Since $ D^{k}_{0q}(\phi,\theta,0)=(-1)^{q}\,\sqrt{\frac{4\pi}{2k+1}}\,Y^{k}_{q}(\theta,\phi) $ $ ^{32} $, 
 \begin{equation}
(t^{k}_{q})^{R}=(-1)^{q}\,\sqrt{\frac{4\pi}{2k+1}}\,t^{k}_{0}\,Y^{k}_{q}(\theta,\phi)
\end{equation}
where $ t^{k}_{0} $ is a real number. It is very well-known$ ^{32} $ that 
 \begin{equation}
 \fl Y^{k}_{q}(\theta,\phi)=\frac{1}{r^{k}}\,\sqrt{\frac{(2k+1)!!}{4\pi k!}}(...((\hat{Q}(\theta,\phi)\otimes\hat{Q}(\theta,\phi))^{2}\otimes\hat{Q}(\theta,\phi))^{3}\otimes...)^{k-1}\otimes\hat{Q}(\theta,\phi))^{k}_{q} \,
 \end{equation}
where $ Y^{k}_{q}(\theta,\phi) $ are the spherical harmonics. Thus every $ t^{k}_{q} $ in the rotated frame is characterized by $ k $ axes $ (k=1,2...2j) $ which are collinear and every axis is given by $ (\theta,\phi) $ and $ (\pi-\theta,\pi+\phi) $. In the unrotated coordinate frame, it is obvious that $ t^{k}_{0}\,{'}s $ are characterized by $ k $ axes collinear to the $ z- $axis. In this case, the MAR consists of $ r_{k} $ spheres and each sphere has $ k $ collinear axes.

\subsubsection{Local Unitary Invariants (LUI)}

   It has been shown $ ^{12} $ that since $ (\hat{Q}(\theta_{i},\phi_{i})\otimes\hat{Q}(\theta_{j},\phi_{j}))^{0}_{0} $ is an invariant under rotation, one can construct in general $ ^{j(2j+1)}C_{2} $ invariants out of $ j(2j+1) $ axes together
with $ 2j $ real positive scalars specifying a spin-$ j $ density matrix. Here $ ^{j(2j+1)}C_{2} $ denotes binomial coefficient. 

 For example, spin-1 or symmetric two qubit state is in general parametrized in terms of 3 axes and 2 real scalars and possess $ ^{3}C_{2} $+2=5 invariants.
spin-3/2 or symmetric three qubit state is represented by 6 axes and 3 real scalars and has $ ^{6}C_{2} $+3 =18
 invariants and 
spin-2 or symmetric four qubit  state is characterized by 10 axes and 4 real scalars and has $ ^{10}C_{2} $+4=49 invariants.

  The importance of LUI in the context of LU classification is brought out in the next section.
\subsection{ Classification of Pure and Mixed States based on Multiaxial Representation}
  Since the most general symmetric state $ \rho(\vec{J}) $ is parametrized in terms of spherical tensors $ t^{k}$ $(k=0,1,...,2j)$ and $ t^{k}\,{'}s $ are characterized by $ k $ axes, it is but natural that the two parameters employed in the entanglement classification of symmetric $ N $-qubit pure states based on MR viz, the degeneracy configuration and the diversity degree, are defined for   $t^{k}\,{'}s $. Thus the degeneracy number here represents the number of identical axes characterizing the given spherical tensor parameter $ t^{k} $. Further, we define the degeneracy configuration $\mathcal{D}_{\{n_{i}\}}$ of $t^k$ as the set of degeneracy numbers  $\{n_{i}\}$ ordered by convention in the decreasing order. The number of $n_{i}\,{'}s$ define the diversity degree of the given $t^k$ and we have $\sum_{i=1}^{k} n_{i}=k $ and $k=1,2,..,2j $. Therefore, the degeneracy configuration here is also the partition of $k$ as introduced in $ ^{14} $ and the number of different configurations is given by the partition function $p(k)$. In addition to this, in the case of symmetric mixed system we need to define another number called the rank $k$ which refers to the rank of the spherical tensor parameter $t^k$. Thus the notation for the degeneracy configuration of $ t^{k} $ becomes $\mathcal{D}^{k}_{\{n_{i}\}}$. Therefore every symmetric $ N $-qubit state is in general characterized by $ N $ configurations. 
  
  For example, in the case of a symmetric two qubit system, the density matrix is characterized by $ t^{1} $
and $ t^{2} $. In the case of $ t^{1} $,
 there is only one axis. Thus  $t^{1}\in \mathcal{D}^{1}_{ 1} $. In the case of $ t^{2} $, there are two axes in general. If the two axes are identical then $t^{2}\in \mathcal{D}^{2}_{ 2} $ and if the axes are not collinear then $t^{2}\in \mathcal{D}^{2}_{ 1,1} $. Thus, in general, a symmetric two-qubit state belongs to either of the two following classes:\\
 $ \{\mathcal{D}^{1}_{ 1},\mathcal{D}^{2}_{ 2}\} $ or  $ \{\mathcal{D}^{1}_{ 1},\mathcal{D}^{2}_{ 1,1}\} $.

 Similarly, a symmetric three qubit system is characterized by $ t^{1} $, $ t^{2} $ and $ t^{3} $. Thus the spin-3/2 or symmetric three qubit density matrix in general belong to one of the following configurations:\\
  $ \{\mathcal{D}^{1}_{ 1},\mathcal{D}^{2}_{ 2}, \mathcal{D}^{3}_{ 3}\} ,   \{\mathcal{D}^{1}_{ 1},\mathcal{D}^{2}_{ 2}, \mathcal{D}^{3}_{ 2,1}\} , \{\mathcal{D}^{1}_{ 1},\mathcal{D}^{2}_{ 2}, \mathcal{D}^{3}_{ 1,1,1}\} ,  \{\mathcal{D}^{1}_{ 1},\mathcal{D}^{2}_{ 1,1}, \mathcal{D}^{3}_{ 3}\} ,  \{\mathcal{D}^{1}_{ 1},\mathcal{D}^{2}_{ 1,1}, \mathcal{D}^{3}_{ 2,1}\} , \\ \{\mathcal{D}^{1}_{ 1},\mathcal{D}^{2}_{ 1,1}, \mathcal{D}^{3}_{ 1,1,1}\} $.\\
 
 Two density matrices are LU equivalent provided they have the same set of invariants. It is obvious from our classification scheme, that density matrices having different sets of degeneracy configurations can never be LU equivalent as they have different set of invariants. Therefore LU equivalent density matrices need to have the same set of degeneracy configurations.

\section{Comparison between Majorana Representation and Multiaxial Representation of the GHZ State} 

 In order to bring out the similarities and the differences between MR and MAR, we take up the $ N $-qubit GHZ state for a detailed investigation. \\
{\bf{MR of GHZ State:}}\\
Consider symmetric $ N $-qubit GHZ state 
 \begin{equation}
 | \psi_{GHZ}\rangle=\frac{1}{\sqrt{2}} \Big[ |\frac{N}{2}\frac{N}{2}\rangle+|\frac{N}{2}\frac{-N}{2}\rangle\Big]\equiv \frac{1}{\sqrt{2}} \Big[ |jj\rangle+|j-j\rangle\Big].
 \end{equation}
 The MR polynomial equations (7) and (9), takes the form, 
 \begin{equation}
 (-1)^{2j}Z^{2j}+1=0
 \end{equation}
Depending on whether $ N $ is odd or even we have the following solutions:
\\\\
Odd $ N$(Half odd integral $ j $) :\,\,\,\,\, $  Z=e^{\frac{2\pi i}{2j}r}$;\,\,\, $\,\,\,  r=0,1,2,...,2j-1$.\\
   Thus the $ 2j $ distinct spinors characterizing $ N $-qubit GHZ state are $ (\frac{\pi}{2},0),\,(\frac{\pi}{2},\frac{2\pi}{2j}),\,(\frac{\pi}{2},\frac{4\pi}{2j})\\,....,(\frac{\pi}{2},\frac{2(2j-1)\pi}{2j}). $
\\\\    
Even $ N$(integral $  j$) :\,\,\,\,\, $  Z=e^{\frac{2\pi i}{2j}(r-\frac{1}{2})}$;\,\,\, $\,\,\,  r=0,1,2,...,2j-1$.
In this case we have $ 2j $ distinct spinors namely $ (\frac{\pi}{2},\frac{\pi}{2j}),\,(\frac{\pi}{2},\frac{3\pi}{2j}),\,(\frac{\pi}{2},\frac{5\pi}{2j}),....,(\frac{\pi}{2},\frac{(4j-1)\pi}{2j}) $ or equivalently $ j $ distinct axes.\\
 According to Bastin et, al.$ ^{14} $, $ N $-qubit GHZ state belong to $ \mathcal{D}^{N}_{\underbrace{1,1,1 \dots 1}_{N}} $ or equivalently $ \mathcal{D}^{2j}_{\underbrace{1,1,1 \dots 1}_{2j}} $ for both odd and even $ N'$s.\\
{\bf{MAR of GHZ State:}}\\
 To find out the axes, consider the density matrix of $ N $-qubit GHZ state in the $ | jm\rangle $ basis; $ m=+j...-j $ 
 \begin{equation}
{\rho_{GHZ} } = \frac{1}{2}\left(
\begin{array}{cccc}
1 & 0 & \ldots & 1 \\
0 &0 & \ldots &0 \\
\vdots & \vdots & \ddots & \vdots \\
1 & 0 & \ldots & 1
\end{array}
\right).
\end{equation} 
\begin{equation}
t^{k}_{q}=\sum_{m=-j}^{+j}\,\rho_{mm^{\prime}}\,[k]\,C(jkj;mqm^{\prime}),\,\,\,\,\,\,\,\,\, with\,\,\,\, m^{\prime}=m+q.
\end{equation} 
Since $ \rho_{jj}=\rho_{j-j}=\rho_{-jj}=\rho_{-j-j}=\frac{1}{2} $ are the only non-zero matrix elements of $ \rho_{GHZ} $ , the $t^{k}_{q}\,{'}s$ can be computed as,   
\begin{equation}
   t^{k}_{q}=0,\,\,\,\,\,\,\,\,\,\,\,   for\,\, all \,\,\, q\neq0,2j. 
  \end{equation}
  Further,  
 \begin{equation}
t^{k}_{0}=\rho_{jj}\,\,[k]\,\,C(jkj;j0j)+\rho_{-j-j}\,\,[k]\,\,C(jkj;-j0-j)=0\,\,\,\,\,\,\, for\,\, odd\,\,k\,{'}s. 
\end{equation} 
 Here we have used the symmetry property of Clebsch-Gordan coefficients namely $ C(jkj;j0j)=(-1)^{k}\,C(jkj;-j0-j) $.
Also
\begin{equation}
t^{k}_{0}=\frac{[k]}{2}\, (2j)!\,\Big[\frac{2j+1}{(2j-k)!\,(2j+k+1)!}\Big]^{1/2},\,\,\,\,\,\,\,\,\,for\,\,\,even\,\,k\,{'}s 
\end{equation}  
since \[ C(cbc;c0c)=(2c)!\,\Bigg[ \frac{(2c+1)}{(2c-b)!(2c+b+1)!}\Bigg]^{1/2} \] (eq. (42) in page 252 of ref.$ ^{32} $).

 To write the polynomial equation for MAR, we compute $ t^{2j}_{\pm 2j} $ as
 \[t^{2j}_{2j}=(-1)^{2j}\,t^{2j}_{-2j}=(-1)^{2j}\,\rho_{-jj}\,\,[2j]\,\,C(j2jj;j-2j-j)\]
 \begin{equation}
 =(-1)^{2j}\,\frac{[2j]}{2}\,\Bigg[ \frac{(2j+1)(4j)!}{(4j+1)!}\Bigg]^{1/2}.\,\,\,\,\, 
 \end{equation}
 Here we have used the expression \[\fl  C(abc;a\beta\gamma)=\delta_{\gamma-\beta,a}\,\Bigg[ \frac{(2c+1)(2a)!(-a+b+c)!(b-\beta)!(c+\gamma)!}{(a+b+c+1)!(a-b+c)!(a+b-c)!(b+\beta)!(c-\gamma)!}\Bigg]^{1/2} \]
  (eq. (36) in page 251 of ref.$^{32}  $).
  
  As in the case of MR, here also we take up the case of odd $ N $ and even $ N $ separately.
 \\\\ 
Odd $ N $ (half odd integral $ j $):
Since $ t^{k}_{0}=0 $ for odd $k\,{'}$s and $ t^{k}_{0}\neq0 $ for even $k\,{'}$s, there exist $ k $ axes collinear to $ z $-axis as explained in sec 3.1 for every even $ k $ $(k=2,4,6 \dots 2j-1) $. Thus, the total number of axes collinear to $ z $-axis, characterizing the odd $ N $-qubit GHZ state is, 
\begin{equation}
2+4+6+...+2j-1=j^{2}-\frac{1}{4}.
\end{equation} 

 Further, for the highest value of $ k $, 
\begin{equation}
P(Z)=\sqrt{^{4j}\,C_{4j}}\,t^{2j}_{2j}\,Z^{0}+\sqrt{^{4j}\,C_{0}}\,t^{2j}_{-2j}\,z^{4j}=0,
\end{equation} 
since $t^{2j}_{ 2j}=-t^{2j}_{-2j}$, we have
 \begin{equation}
P(Z)=Z^{4j}-1=0,
\end{equation}

\begin{equation}
Z=e^{\frac{2\pi i}{4j}r},\,\,\,\,\,\,\,\,\,\,\,\, r=0,1 \dots 4j-1.
\end{equation}
There exist $ 4j $ solutions or $ 2j $  axes namely 
\begin{equation}
(\frac{\pi}{2},0),\,(\frac{\pi}{2},\frac{\pi}{2j}),\,(\frac{\pi}{2},\frac{2\pi}{2j}) \dots (\frac{\pi}{2},\frac{(4j-1)\pi}{2j}).
\end{equation}

Therefore, the degeneracy configuration of the statistical tensor parameters are given by
\begin{equation}
t^{2}\in \mathcal{D}^{2}_{2},\,t^{4}\in \mathcal{D}^{4}_{4} \dots t^{2j-1}\in \mathcal{D}^{2j-1}_{2j-1},\,t^{2j}\in \mathcal{D}^{2j}_{\underbrace{1,1,1 \dots 1}_{2j}} 
\end{equation}
Thus according to our classification, the degeneracy configuration of $ N $-qubit GHZ state for odd $ N $ is
$ \{\mathcal{D}^{2}_{2},\mathcal{D}^{4}_{4},\mathcal{D}^{6}_{6},...,\mathcal{D}^{2j-1}_{2j-1},\mathcal{D}^{2j}_{\underbrace{1,1,1 \dots 1}_{2j}} \}$.
\\\\
Even $ N $ (integral $ j $): Since $ t^{k}_{0}\neq0 $ for $ k=2,4,6,...,2j-2 $, there exist $ k $ axes collinear to $ z $-axis. Thus, in this case the total number of axes collinear to the $ z $-axis is,
\begin{equation}
2+4+6+...+2j-2=j(j-1)
\end{equation}
The polynomial equation for the highest $ k $ is,
\begin{equation}
P(Z)=\sqrt{^{4j}\,C_{2j}}\,t^{2j}_{0}\,Z^{2j}+\sqrt{^{4j}\,C_{4j}}\,t^{2j}_{2j}\,Z^{0}+\sqrt{^{4j}\,C_{0}}\,t^{2j}_{-2j}\,Z^{4j}=0.
\end{equation} 
Since in this case  $t^{2j}_{ 2j}=t^{2j}_{-2j}$, we have
 \begin{equation}
P(Z)=\sqrt{^{4j}\,C_{2j}}\,t^{2j}_{0}\,Z^{2j}+t^{2j}_{2j}\,\Big(Z^{4j}+1 \Big)=0.
\end{equation} 
Substituting $t^{2j}_{ 0}$ and $t^{2j}_{ 2j}$ from eq. (38) and eq. (39) respectively, 
\begin{equation}
\fl P(Z)=\Big[\frac{(4j)!}{(2j)!\,(2j)!}\Big]^{1/2}\,2\,(2j)!\,\Big[\frac{2j+1}{(4j+1)!}\Big]^{1/2}\,Z^{2j}+\Big[\frac{(4j)!\,(2j+1)}{(4j+1)!}\Big]^{1/2}\Big(Z^{4j}+1\Big) =0
\end{equation}
which leads to
\begin{equation}
P(z)=Z^{4j}+2Z^{2j}+1 =0.
\end{equation}
Thus
\[(Z^{2j}+1)^{2}=0,
\]
\begin{equation}
Z=e^{\frac{2\pi i}{2j}(r-\frac{1}{2})},\,\,\, r=0,1,...,2j-1
\end{equation}
There exist two identical sets of solutions or $j $ axes namely
\begin{equation}
(\frac{\pi}{2},\frac{\pi}{2j}),\,(\frac{\pi}{2},\frac{3\pi}{2j}),\,(\frac{\pi}{2},\frac{5\pi}{2j})...(\frac{\pi}{2},\frac{(4j-1)\pi}{2j}).
\end{equation}
Therefore, the degeneracy configuration of the statistical tensor parameters are given by
\begin{equation}
t^{2}\in \mathcal{D}^{2}_{2},\,\,t^{4}\in \mathcal{D}^{4}_{4} \dots t^{2j-2}\in \mathcal{D}^{2j-2}_{2j-2},\,\,t^{2j}\in \mathcal{D}^{2j}_{\underbrace{2,2 \dots 2}_{j}} 
\end{equation}
Thus according to our classification the degeneracy configuration of $ N $-qubit GHZ state for even $ N $ is
$  \{\mathcal{D}^{2}_{2},\,\mathcal{D}^{4}_{4},\,\mathcal{D}^{6}_{6},...,\mathcal{D}^{2j-2}_{2j-2},\,\mathcal{D}^{2j}_{\underbrace{2,2 \dots 2}_{j}} \}$.

 Let us now consider The MR and MAR of the 3-qubit and 4-qubit GHZ states.
\\\\
MR of 3-qubit GHZ state:

   Consider  $|\psi_{GHZ}\rangle = \frac{|\frac{3}{2},\frac{3}{2}\rangle + |\frac{3}{2},-\frac{3}{2}\rangle}{\sqrt2}  \equiv  \frac{|\uparrow\uparrow\uparrow\rangle+|\downarrow\downarrow\downarrow\rangle}{\sqrt 2}.$ \\
Since $  N$ is odd, according to eq. (33), the polynomial equation is given by $ Z^{3}=1 $ and the three distinct spinors are, 
\begin{equation}
(\frac{\pi}{2},0),\, (\frac{\pi}{2},\frac{2\pi}{3}),\,(\frac{\pi}{2},\frac{4\pi}{3}).\\
 \end{equation}
 Thus, $ |\psi_{GHZ}\rangle\in \mathcal{D}^{3}_{1,1,1} $.\\
 Spinors characterizing the MR of the 3-qubit GHZ state are shown in figure 1.
\begin{figure}[h]
\begin{center}
\includegraphics[width=4.2cm]{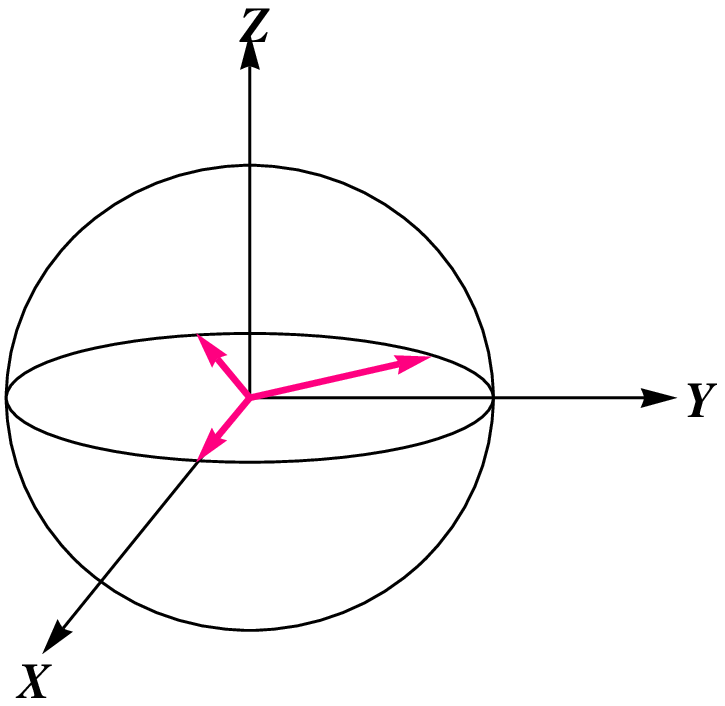}
\vspace{-5mm}
\end{center}
\caption{(color online)MR of the 3-qubit GHZ state.}
\end{figure} 
 \newpage  
MAR of 3-qubit GHZ state:

   Corresponding density matrix for 3-qubit GHZ state is
\begin{equation}
\rho_{GHZ}=\frac{1}{2}
\left(\begin{array}{ccccc}
1 & 0 & 0 & 1\\

0 & 0 & 0 & 0 \\
 
0 & 0 & 0 & 0 \\

1 & 0 & 0 & 1
\end{array}\right)\,.
\end{equation}
The non-zero $t^{k}_{q}\,{'}s$ from eq. (17) are:
\begin{equation}
t^{2}_{0}=\rho_{\frac{3}{2},\frac{3}{2}}\,\sqrt{5}\,C(\frac{3}{2}2\frac{3}{2};\frac{3}{2}0\frac{3}{2})+\rho_{\frac{-3}{2},\frac{-3}{2}}\,\sqrt{5}\,C(\frac{3}{2}2\frac{3}{2};\frac{-3}{2}0\frac{-3}{2})=1
\end{equation}
\begin{equation}
t^{3}_{3}=\rho_{\frac{-3}{2},\frac{3}{2}}\,\sqrt{7}\,C(\frac{3}{2}3\frac{3}{2};\frac{-3}{2}3\frac{3}{2})=-1
\end{equation}
\begin{equation}
t^{3}_{-3}=\rho_{\frac{3}{2},\frac{-3}{2}}\,\sqrt{7}\,C(\frac{3}{2}3\frac{3}{2};\frac{3}{2}-3\frac{-3}{2})=1.
\end{equation}
 Solving the polynomial equation for $ t^{3}_{q}; $ $ q=3,-3 $ (eq. (41)), we have
\begin{equation}
Z=e^{\frac{2\pi i}{6}r},\,\,\,\,\,\,\, r=0,1,...,5
\end{equation}
Thus, the three distinct axes are:
\begin{equation}
(\frac{\pi}{2},0),\, (\frac{\pi}{2},\pi),\,(\frac{\pi}{2},\frac{\pi}{3}),\, (\frac{\pi}{2},\frac{4\pi}{3}),\,(\frac{\pi}{2},\frac{2\pi}{3}),\,(\frac{\pi}{2},\frac{5\pi}{3}).
\end{equation}
Also, since $ t^{2}_{0}=1 $, there exist two axes collinear to $ z $-axis.\\
Therefore, $ t^{2}\in \mathcal{D}^{2}_{2} $, $ t^{3}\in \mathcal{D}^{3}_{1,1,1} $ and $ \rho\in \{\mathcal{D}^{2}_{2},\,\mathcal{D}^{3}_{1,1,1}\} $.\\
Axes characterizing MAR of 3-qubit GHZ state are shown in figure 2.
\begin{figure}[h]
\subfloat[$ t^{2}\in\mathcal{D}^{2} _{2}$
\label{fig:test1}]
  {\includegraphics[width=.33\linewidth]{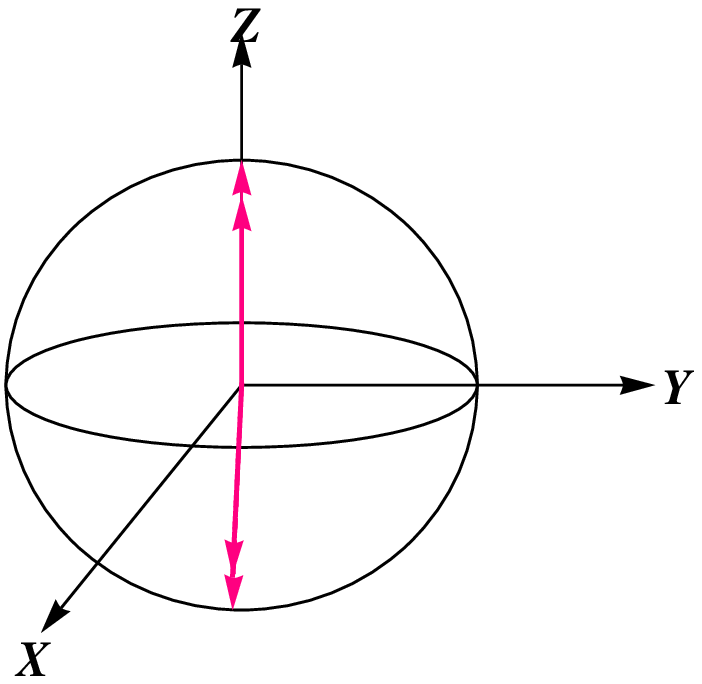}}\hfill
\subfloat[$ t^{3}\in\mathcal{D}^{3} _{1,1,1}$
\label{fig:test2}]
  {\includegraphics[width=.4\linewidth]{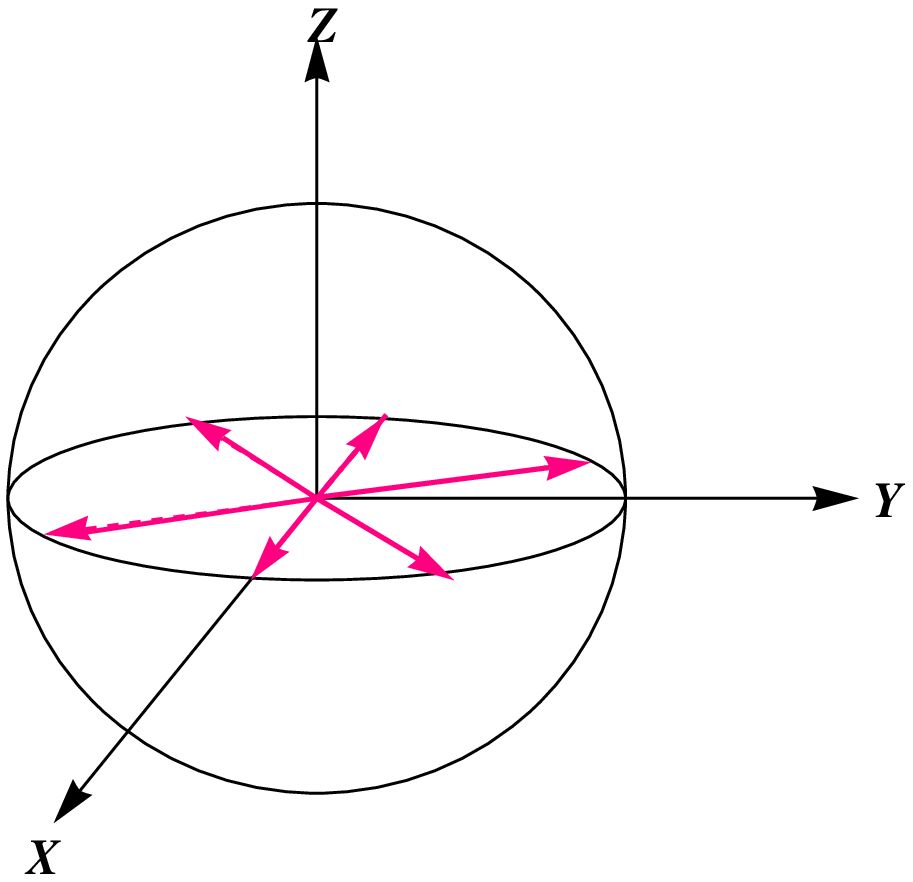}}\hfill
\caption{ (color online) MAR of   $ t^{2} $ and $ t^{3} $ characterizing the 3-qubit GHZ state.}
\end{figure}
\newpage
MR of 4-qubit GHZ state:

 Corresponding 4-qubit GHZ state in $|2m\rangle $ representation is, $|\psi_{GHZ}\rangle = \frac{|2,2\rangle + |2,-2\rangle}{\sqrt2}.$ \\
Since $  N$ is even, the polynomial equation is given by $ Z^{4}=-1 $ which leads to
\begin{equation}
Z=e^{\frac{2\pi i}{4}(r-\frac{1}{2})},\,\,\,\,\,\,\,\,\, r=0,1,2,3
\end{equation}
 We get four distinct spinors or equivalently two distinct axes 
\begin{equation}
(\frac{\pi}{2},\frac{\pi}{4}),\, (\frac{\pi}{2},\frac{3\pi}{4}),\,(\frac{\pi}{2},\frac{5\pi}{4}),\,(\frac{\pi}{2},\frac{7\pi}{4})
\end{equation}
Thus, $ |\psi_{GHZ}\rangle\in \mathcal{D}^{4}_{1,1,1,1} $.\\
Spinors characterizing MR of 4-qubit GHZ state are shown in figure 3.

 \begin{figure}[h]
\begin{center}
\includegraphics[width=5.1cm]{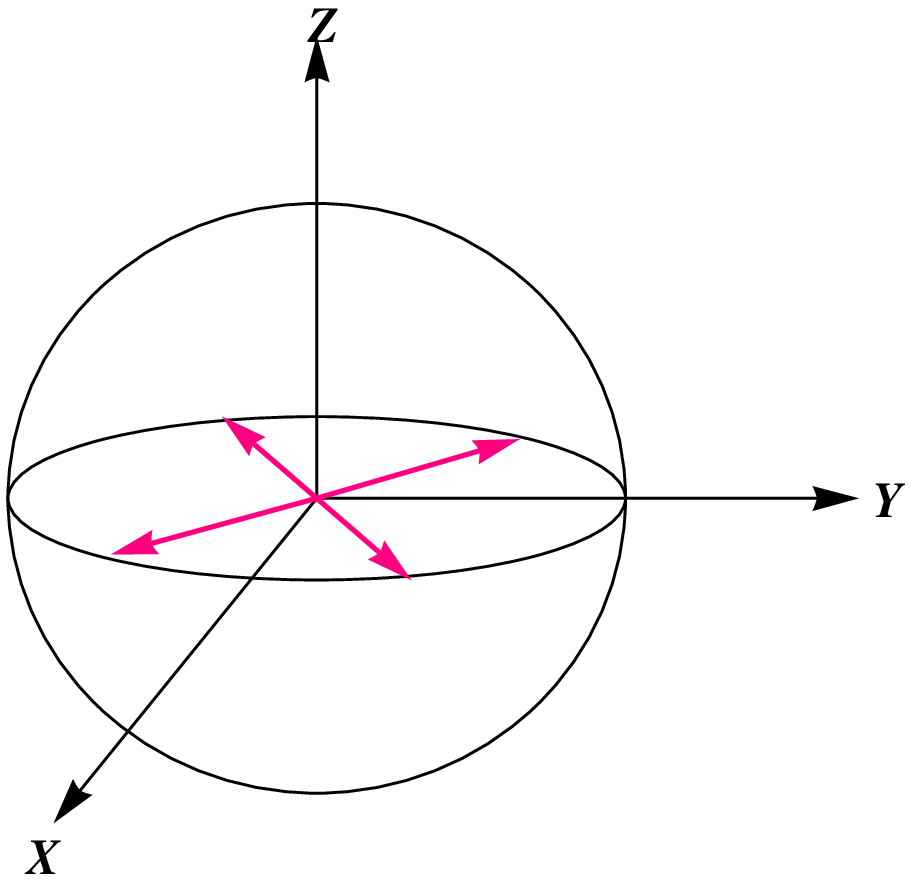}
\vspace{-5mm}
\end{center}
\caption{(color online)MR of the 4-qubit GHZ state.}
\end{figure}

\newpage
MAR of 4-qubit GHZ state:

    Corresponding density matrix for 4-qubit GHZ state is
\begin{equation}
\rho_{GHZ}=\frac{1}{2}
\left(\begin{array}{ccccc}
1 & 0 & 0 &0 & 1\\

0 & 0 & 0 &0 & 0 \\
 
0 & 0 & 0 &0 & 0 \\
0 & 0 & 0 &0 & 0 \\
1 & 0 & 0 &0 & 1
\end{array}\right)\,.
\end{equation}
The only non-zero $t^{k}_{q}\,{'}s$ are:
\begin{equation}
t^{2}_{0}=\sqrt{\frac{10}{7}},\,\,\,t^{4}_{0}=\frac{1}{\sqrt{14}},\,\,\,
t^{4}_{4}=\frac{\sqrt{5}}{2},\,\,\,t^{4}_{-4}=\frac{\sqrt{5}}{2}.
\end{equation}
Since $ N$ is even, solving the polynomial equation for $ t^{4}_{q}; $ $ q=0,4,-4 $ we get 
\begin{equation}
Z=e^{\frac{2\pi i}{4}(r-\frac{1}{2})},\,\,\,\,\,\,\, r=0,1,2,3
\end{equation}
Thus we get two sets of two distinct axes and the axes are given by
\begin{equation}
(\frac{\pi}{2},\frac{\pi}{4}),\, (\frac{\pi}{2},\frac{3\pi}{4}),\,(\frac{\pi}{2},\frac{5\pi}{4}),\,(\frac{\pi}{2},\frac{7\pi}{4}).
\end{equation}
In this case $ t^{2}\in \mathcal{D}^{2}_{2} $ and $ t^{4}\in \mathcal{D}^{4}_{2,2} $, thus $ \rho\in \{\mathcal{D}^{2}_{2},\,\mathcal{D}^{4}_{2,2}\} $.\\
Axes characterizing MAR of 4-qubit GHZ state are shown in figure 4.
\begin{figure}[h]
\subfloat[$ t^{2}\in\mathcal{D}^{2} _{2}$
\label{fig:test1}]
  {\includegraphics[width=.33\linewidth]{fig2}}\hfill
\subfloat[$ t^{4}\in\mathcal{D}^{4} _{2,2}$
\label{fig:test2}]
  {\includegraphics[width=.4\linewidth]{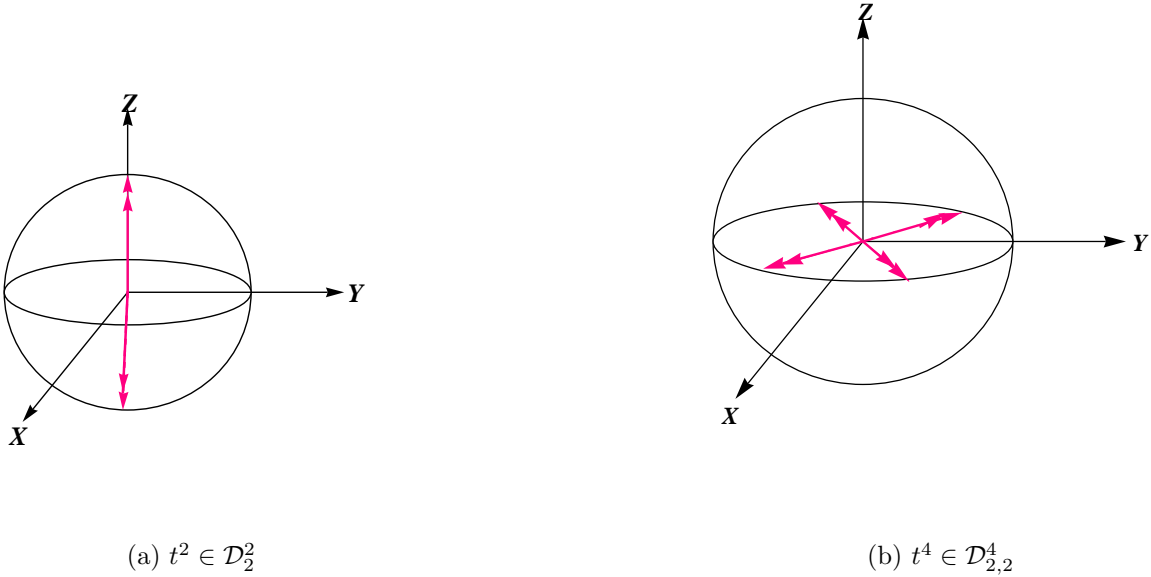}}\hfill
\caption{ (color online) MAR of   $ t^{2} $ and $ t^{4} $ characterizing the 4-qubit GHZ state.}
\end{figure}

Thus it is evident that MR is not a special case of MAR. One can also note that the basic entity characterizing MR is a spinor and MAR is an axis.

\section{Classification of Some Well-Known States}
 \subsection{$ N $-Qubit Separable States}
 Let us now consider symmetric $N$-qubit pure separable state $ | \psi^{j}\rangle= |\epsilon,\epsilon,...,\epsilon\rangle $. Following Bastin et. al $ ^{14} $, this state belongs to $ S $ family with diversity degree $ d=1 $. This state has a canonical form $|{S_{n}}\rangle \equiv|\uparrow\uparrow\uparrow...\uparrow\rangle\equiv|jj\rangle$ brought out by a rotation. In other words $\langle{jj}|\rho|{jj}\rangle=1$ in some rotated frame of reference. In the density matrix language, the corresponding density matrix is given by
\begin{equation}
{\rho } = \left(
\begin{array}{cccc}
1 & 0 & \ldots & 0 \\
0 &0 & \ldots &0 \\
\vdots & \vdots & \ddots & \vdots \\
0 & 0 & \ldots & 0
\end{array}
\right),
\end{equation}
in the $|{jm}\rangle$ basis. From eq. (17), the only non-zero spherical tensor parameters characterizing the above state are, 
\begin{equation}
t^{k}_{0}=[k]\,\rho_{jj}\,C(jkj;j0j),\,\,\,\,\,\,\,\,\,\,\, k=0,1,2...2j.
\end{equation} 
 Thus, $ N $-qubit pure separable state is characterized by $ j(2j+1) $ axes collinear to $ z $-axis. Equivalently, in other frames of reference the state has to be constructed out of $j(2j+1)$ axes which are collinear and parallel to the rotated $ z $-axis. Therefore the degeneracy configuration of $ t^{1},t^{2},...,t^{k} $; $ k=1,2,...,2j $ of $\rho$ must be $\left\lbrace{\mathcal{D}^{1}_{1},\mathcal{D}^{2}_{2},\mathcal{D}^{3}_{3},...\mathcal{D}^{k}_{k}}\right\rbrace$ respectively with the understanding that all the axes are collinear. Further, Since $Tr(\rho^{2})=1$ we have from eq. (12) the constraint, 
\begin{equation}
\frac {Tr}{(2j+1)^{2}}\left[ \sum_{kq}t^{k}_{q}\tau^{k^{\dagger}}_{q}\, \sum_{k^{\prime}q^{\prime}}t^{k^{\prime^{*}}}_{q^{\prime}}\tau^{k^{\prime}}_{q^{\prime}}\right]
=\frac {1}{(2j+1)^{2}}\sum_{kqk^{\prime}q^{\prime}}t^{k}_{q}t^{k^{\prime^{*}}}_{q^{\prime}}\,Tr(\tau^{k^{\dagger}}_{q}\tau^{k^{\prime}}_{q^{\prime}}) 
\end{equation}
Using (14) and (24) we obtain,
\[\frac {1}{(2j+1)} \sum^{2j}_{k=0}t^{k}.t^{k}= 1\,\,\Rightarrow \frac {1}{(2j+1)} \sum^{2j}_{k=0}r^{2}_{k}(\hat{Q}(\theta,\varphi)\otimes\hat{Q}(\theta,\varphi)\otimes.......\,\]
 
 \begin{equation}
  \otimes\hat{Q}(\theta,\varphi))^{k}.(\hat{Q}(\theta,\varphi)\otimes\hat{Q}(\theta,\varphi)\otimes.......\otimes\hat{Q}(\theta,\varphi))^{k}= 1,
  \end{equation}
where $ t^{k}.t^{k}=\sum_{q}(-1)^{q}t^{k}_{-q}t^{k}_{q} $. Also, from eq. (24), we can calculate the local unitary invariants (LUI) $r_{k}\,{'}s$ $ ^{12} $ as,
\begin{eqnarray*}
\fl  r_{k}=\frac{t^{k}_{0}}{(\hat{Q}(\theta,\varphi)\otimes\hat{Q}(\theta,\varphi)\otimes...\hat{Q}(\theta,\varphi))^{k}_{0}}
= \frac{[k]\, C(jkj;j0j)}{C(112;000).\,C(213;000)...C(k-11k;000)}\nonumber\\
\end{eqnarray*}
\begin{eqnarray}
 r_{k}&=&\frac{[k]\,(2j)!\,\left(\frac{2j+1}{(2j-1)!(2j+k+1)!}\right)^\frac{1}{2}}{\prod_{n=1,.,k}\frac{n!}{(n-1)!}\left(\frac{2!\,2(n-1)!}{(2n)!}\right)^\frac{1}{2}},
\end{eqnarray}
using the expression for Clebsch Gordan coefficients (from eq. (33) and eq. (42) in pages 251 and 252 of ref.$ ^{32} $).

 For example for a two qubit pure separable state,
\begin{equation}
{\rho}=
\left(\begin{array}{cccc}
1 & 0 & 0\\

 0 & 0 & 0 \\

 0 & 0 & 0
\end{array}\right)\,.
\end{equation} 
and the corresponding $r_{k}\,{'}s$ are: 
\begin{equation}
 r_{1}=\sqrt{\frac{3}{2}},\,\,\,\,\, r_{2}={\frac{\sqrt{3}}{2}}.
\end{equation}
 For three qubit pure separable state,
\begin{equation}
{\rho}=
\left(\begin{array}{cccc}
1 & 0 & 0& 0\\

 0 & 0 & 0& 0 \\

 0 & 0 & 0& 0\\
  0 & 0 & 0& 0
\end{array}\right)\,.
\end{equation}  
 and the corresponding $r_{k}\,{'}s$ are: 
 \begin{equation}
  r_{1}={\frac{3}{\sqrt 5}},\,\,\,\, r_{2}={\sqrt\frac{{3}}{2}},\,\,\,\,  r_{3}={\frac{{1}}{\sqrt 2}}.
 \end{equation} 
 Thus to check for the separability of a given pure state the recipe is as follows: compute $Tr(\rho^{2})$. If $Tr(\rho^{2})=1$, compute all the $t^{k}_{q}\,{'}s$. Then solve the polynomial equations (23) for each $t^{k}_{q}\,{'}s$ and obtain the axes. Even if one of the axes is different from the rest, the state is not separable. Thus, the necessary but not a sufficient condition for a pure state to be separable is that all the axes are collinear. If all the axes are collinear, then compute the values of $r_{k}\,{'}s$. The given state is separable iff all the $r_{k}\,{'}s$ so obtained are equal to those given by eq. (71). If $Tr(\rho^{2})<1$, it is not clear as to the procedure to be followed to test the separability of a given $N$-qubit state as the definition of entanglement for a mixed state itself is problematic. Before we look into the problem of mixed state entanglement, let us now consider some well known examples of two qubit and three qubit entangled pure states and $N$-qubit GHZ state to demonstrate our method.\\
\subsection{Pure Entangled States}
 \subsubsection{{\bf Bell State}}
Consider one of the Bell states  $|\psi\rangle = |10\rangle \equiv \frac{|\uparrow\downarrow\rangle+|\downarrow\uparrow\rangle}{\sqrt 2}$ which is a symmetric state. The corresponding density matrix in $ |1m\rangle $ basis; $ m=1,0,-1 $ is,  
\begin{equation}
{\rho}=\frac{1}{2}
\left(\begin{array}{cccc}
0 & 0 & 0\\

 0 & 1 & 0 \\

 0 & 0 & 0
\end{array}\right)\,.
\end{equation} 
The only non-zero $t_{q}^{k}$ from eq. (17) is
\begin{equation}
t^{2}_{0} = \sqrt{ 2}.
\end{equation}
Solving the polynomial equation for $t^{2}$, we get $ Z^{2}=0 $ which show that the two axes are collinear to $ z $-axis.
Thus $t^{2}$ belongs to $\mathcal{D}^{2}_{2}$ but $r_{2}=\sqrt{3}$ and according to eq. (71), the Bell state is not separable. Axes and the invariants characterizing the Bell state are shown in figure 5. Therefore according to MAR, $ \rho\in\{\mathcal{D}^{2}_{2}\} $ in contrast to the degeneracy configuration of the Bell state based on MR namely $ \{\mathcal{D}^{2}_{1,1}\} $. 
 
\begin{figure}[h]
\begin{center}
\includegraphics[width=4.0cm]{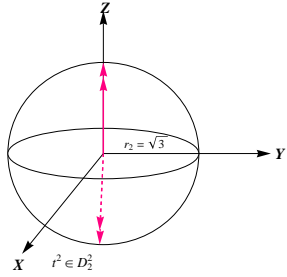}
\vspace{-5mm}
\end{center}
\caption{(color online) MAR of the Bell state. $\psi\rangle =  \frac{|\uparrow\downarrow\rangle+|\downarrow\uparrow\rangle}{\sqrt 2}$
 The two axes representing the Bell state are along the z-axis and shown by double headed arrows.}
\end{figure} 

  \subsubsection{{\bf W state}}  
 
 Consider the symmetric three qubit state, $|\psi_{W}\rangle \equiv\frac{|\uparrow\downarrow\downarrow\rangle+ |\downarrow\uparrow\downarrow\rangle+|\downarrow\downarrow\uparrow\rangle}{\sqrt{3}}= |3/2~ -1/2\rangle$.
The corresponding density matrix in $ |\frac{3}{2}m\rangle $ basis; $ m=\frac{3}{2},\frac{1}{2},\frac{-1}{2},\frac{-3}{2} $ is,

\begin{eqnarray}
\rho_{W}= \left(\begin{array}{c}
0~ ~~ 0 ~~~ 0 ~~ 0  \\

0~ ~~ 0 ~~~ 0 ~~ 0  \\

0~ ~~ 0 ~~~ 1 ~~ 0  \\

0 ~~ 0 ~~~ 0 ~~ 0
\end{array}\right)
\end{eqnarray}

The non-zero $t^{k}_{q}{'} s$ from eq. (17) are 
\begin{equation}
t^{1}_{0} = \frac{-1}{\sqrt{5}},\,\,\,\,\,\,  t^{2}_{0} = -1,\,\,\,\,\,\,  t^{3}_{0} = \frac{3}{\sqrt{5}}. 
\end{equation}

Thus all the 6 axes are collinear and parallel to $z$-axis. Thus $t^{1}\in \mathcal{D}^{1}_{1}$, $t^{2}\in \mathcal{D}^{2}_{2}$ and $t^{3}\in \mathcal{D}^{3}_{3}$ and hence $ \rho\in\{\mathcal{D}^{1}_{1},\mathcal{D}^{2}_{2},\mathcal{D}^{3}_{3}\}. $ Here $
r_{1}={\frac{1}{\sqrt{2}}},\,\, r_{2} = \sqrt{\frac{3}{2}},\,\, r_{3} = {\frac{3}{\sqrt{5}}}$ and since $r_{1}$ and $r_{2}$ are not equal to the values obtained from eq. (71), W state is not separable. Axes and the invariants characterizing the W state are shown in figure 6. \\
 Observe that $ |\psi_{W^{*}}\rangle \equiv\frac{|\uparrow\uparrow\downarrow\rangle+ |\downarrow\uparrow\uparrow\rangle+|\uparrow\downarrow\uparrow\rangle}{\sqrt{3}}= |3/2~ 1/2\rangle$ is also represented by the same set of axes and the invariants. Thus MAR does not distinguish between $ |\psi_{W}\rangle $ and $ |\psi_{W^{*}}\rangle $. 

\begin{figure}[h]
\subfloat[$ t^{1}\in\mathcal{D}^{1} _{1}$
\label{fig:test1}]
  {\includegraphics[width=.24\linewidth]{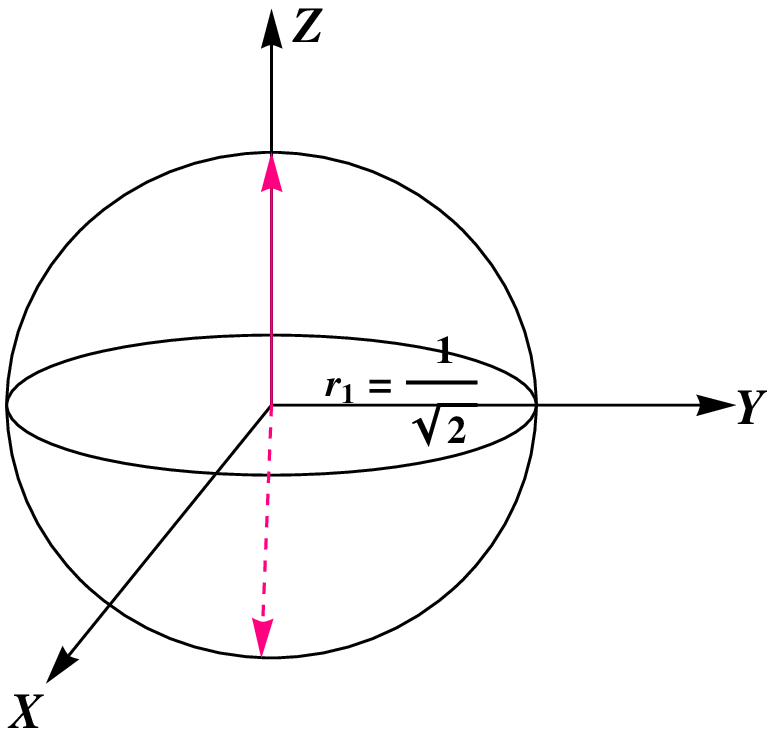}}\hfill
\subfloat[$ t^{2}\in\mathcal{D}^{2} _{2}$
\label{fig:test2}]
  {\includegraphics[width=.30\linewidth]{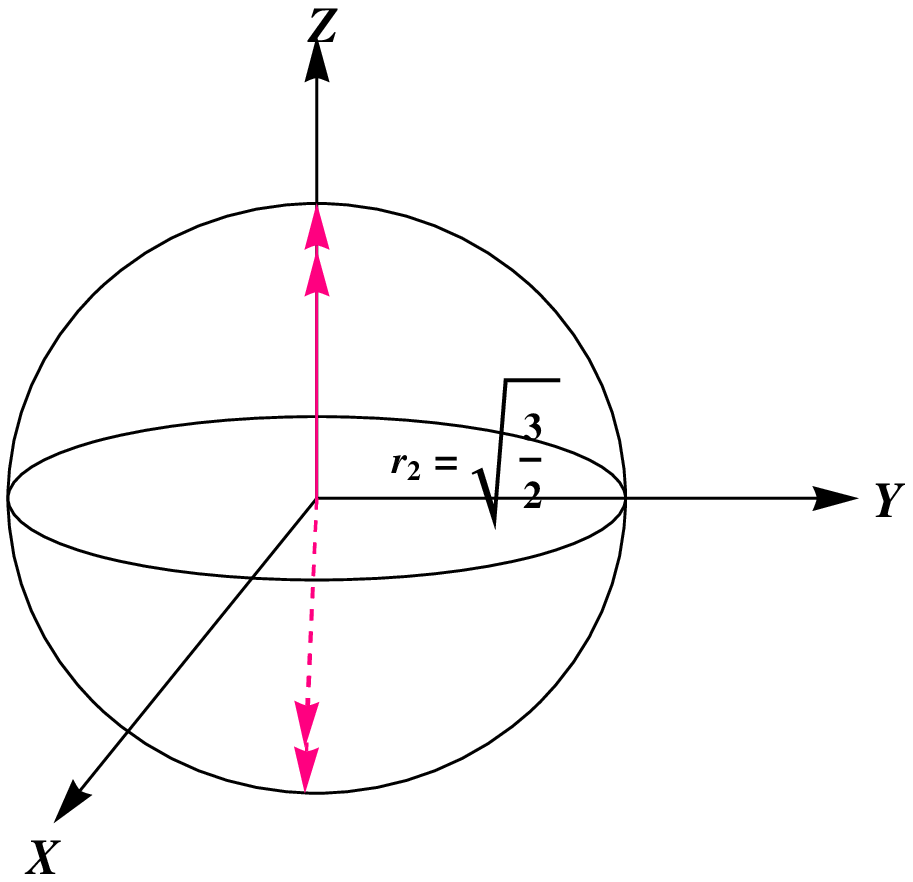}}\hfill
\subfloat[$ t^{3}\in\mathcal{D}^{3} _{3}$
\label{fig:test3}]
  {\includegraphics[width=.36\linewidth]{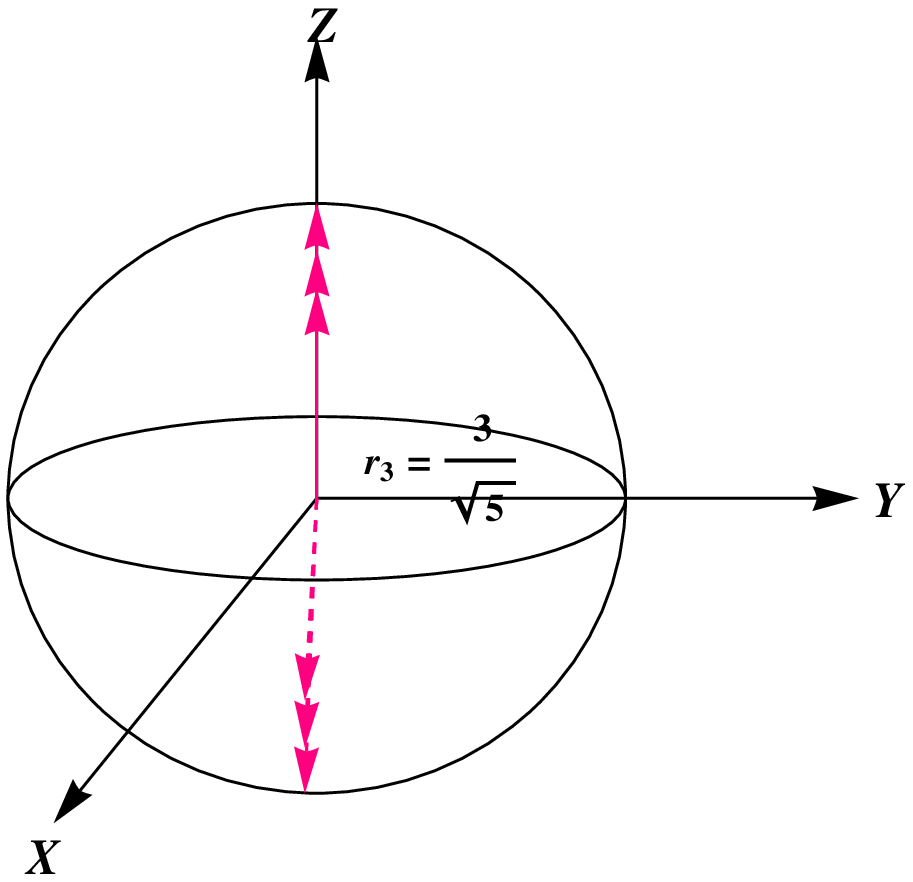}}
\caption{ (color online) Multiaxial representation of  $ t^{1}  $, $ t^{2} $ and $ t^{3} $ characterizing the W state.}
\end{figure}
\newpage
\subsubsection{{\bf $ N $-Qubit GHZ state}}
 Since the degeneracy configuration of $ N $-qubit GHZ state in the case of odd $ N $ is $ \{\mathcal{D}^{2}_{2},\mathcal{D}^{4}_{4},\mathcal{D}^{6}_{6},...,\mathcal{D}^{2j-1}_{2j-1},\mathcal{D}^{2j}_{\underbrace{1,1,1 \dots 1}_{2j}} \}$ and in the case of even $ N $ is $  \{\mathcal{D}^{2}_{2},\,\mathcal{D}^{4}_{4},\,\mathcal{D}^{6}_{6},...,\mathcal{D}^{2j-2}_{2j-2},\,\mathcal{D}^{2j}_{\underbrace{2,2 \dots 2}_{j}} \}$, it is obvious that the axes characterizing the GHZ state are not all collinear and hence the GHZ state is not separable.

\subsection{Mixed States} 
 Let us now consider density matrices representing some well-known spin-1 systems belonging to different LU classes and study their classification in terms of LUI's $ r_{1},\, r_{2} $ and the axes. These density matrices also represent symmetric two qubit mixed states whose entanglement can be studied in terms of the $ r_{k}'$s $ (k=1,2) $ and the axes. Here we make use of the Positive Partial Transpose (PPT) criterion $ ^{33} $ to characterize entanglement.
 
\subsubsection{{Uniaxial Systems }}. 
  
  The density matrix representing a uniaxial system is characterized by $t^{1}_{0} \neq 0,\,t^{1}_{\pm 1} \neq 0,\,\, t^{2}_{q} = 0\,(q=0,\pm1,\pm2)$. In the context of nuclear physics, such a system is said to be purely vector polarized with single axis of cylindrical symmetry. It can be produced in the laboratory by the interaction of a spin-1 assembly with an external dipole magnetic field $ ^{34} $. Here 
The most general density matrix corresponding to this class in the $|1m\rangle$ basis is
\begin{equation}
\rho=\frac{1}{3}
\left(\begin{array}{ccccc}
1+\sqrt {\frac{3}{2}}\,r_{1} cos\theta_{1} & {\frac{\sqrt 3}{2}}r_{1} sin\theta_{1}e^{-i\varphi_{1}} & 0  \\
 {\frac{\sqrt 3}{2}}r_{1} sin\theta_{1}e^{i\varphi_{1}} & 1 & {\frac{\sqrt 3}{2}}r_{1} sin\theta_{1}e^{-i\varphi_{1}}   \\
0 &  {\frac{\sqrt 3}{2}}r_{1} sin\theta_{1}e^{i\varphi_{1}} &  1-\sqrt {\frac{3}{2}}\,r_{1} cos\theta_{1} \\
\end{array}\right)\,.
\end{equation}
The non-zero $t^{k}_{q}\,{'}s$ are
\[t^{1}_{0} =r_{1}\,cos\theta_{1}\,\,,\,\,
t^{1}_{\pm 1} = \frac{r_{1}}{\sqrt{2}}\, sin\theta_{1}e^{\mp i\varphi_{1}}\,\,.\] and the only LU invariant is $ r_{1} $.

 $\rho $ is found to be positive semi-definite iff  $0<r_{1}\leq \sqrt \frac{2}{3}$ and entangled for  $\frac {1}{\sqrt 2}\leq r_{1}\leq \sqrt \frac{2}{3}$ for all values of $ \theta $ ($ 0\leq \theta \leq \pi $). Here $ Tr(\rho^{2})=\frac{1}{3}[1+r_{1}^{2}]<1 $, and hence this class consists of mixed states only.
 
  Since $ t^{1}\in  \mathcal{D}^{1}_{1}, $  $ \rho\in  \{\mathcal{D}^{1}_{1}\}. $

\subsubsection{{ Biaxial Systems }}

   Here $t^{1}_{q}= 0\,(q=0,\pm1),\,\,\, t^{2}_{0}\neq $ 0,\, $t^{2}_{\pm 2}\neq $ 0, $ t^{2}_{\pm 1}= $ 0  
and the two axes which characterize $ \rho $ in the Principal Axes of Alignment frame(PAAF) $ ^{35} $ are $\{(\theta,0),(\pi-\theta,\pi)\} $ and  $ \{(\theta,\pi),(\pi-\theta,0)\}$. Thus $ \rho $ is said to be Biaxial. Since $ \rho $ in this case is parametrized in terms of second rank tensor parameters only, namely $ t^{2}_{0} $ and $ t^{2}_{\pm 2} $, it is said to be purely tensor polarized. Such a system can be produced by the interaction of a spin-1 nuclei with an external electric quadrupole field $ ^{36} $. The density matrix in PAAF is given by \\
\begin{equation}
\rho=\frac{1}{3}
\left(\begin{array}{ccccc}
1+ \frac{1}{2\sqrt{3}}\,r_{2}(1+cos^{2}\theta) & 0 & \frac{-\sqrt{3}}{2}r_{2}sin^{2}\theta \\ 0 & 1-\frac{1}{\sqrt{3}}\,r_{2}(1+cos^{2}\theta) & 0 \\ \frac{-\sqrt{3}}{2}r_{2}sin^{2}\theta & 0 & 1+ \frac{1}{2\sqrt{3}}\,r_{2}
\end{array}\right)\,
\end{equation}

 in $|1m\rangle$ basis.\\ 
 $\rho $ is positive semi-definite iff  $0<r_{2}\leq \sqrt{3}$ and the range of $\theta$ then depends on $r_{2}$.\\
 The non-zero $t^{k}_{q}\,{'}s$ are
\[t^{2}_{0} =\frac{r_{2}}{\sqrt 6}\,(1+cos^{2}\theta)\,\,,\,\,
t^{2}_{\pm 2} = \frac{-r_{2}}{2}\, sin^{2}\theta\,\,.\]
   For $ 0<\theta< \frac{\pi}{2} $ and $ \frac{\pi}{2}<\theta<\pi  $, $t^{2} \in \mathcal{D}^{2}_{1,1}$ and hence $\rho \in \{\mathcal{D}^{2}_{1,1}\}$.\\
 For $\theta=0,\frac{\pi}{2},\pi$, $t^{2} \in \mathcal{D}^{2}_{2}$ and hence $\rho \in \{\mathcal{D}^{2}_{2}\}$.

  $\rho $ is positive semi-definite and separable iff  $0<r_{2}\leq \frac {\sqrt{3}}{4}$ and  $0\leq\theta\leq \pi$.
    For $r_{2}=\sqrt 3 $ and $\theta=\frac{\pi}{2}$, $ \rho $ is pure as well as entangled.\\ 
  
  \subsubsection{{Triaxial Systems}} 
  
   Here $t^{1}_{0}\neq 0,\,\, t^{1}_{\pm 1}= 0,\,\,t^{2}_{0}\neq 0,\,\, t^{2}_{\pm 2}\neq 0,\,\, t^{2}_{\pm 1}= 0   $. Such a system is realized when a spin-1 nucleus with non-zero quadrupole moment is exposed to a combined external dipole and electric quadrupole field found in suitable crystal lattice$ ^{37} $. Consider a special case of the density matrix belonging to this class such that the axes are $ \{(\theta_{1}=0),(\theta_{1}=\pi)\}$, $\{(\theta,0),(\pi-\theta,\pi)\} $ and  $ \{(\theta,\pi),(\pi-\theta,0)\}$. Such a $ \rho $ is called triaxial. The density matrix corresponding to this class is explicitly given by
 \small
 \begin{equation}
\fl \rho=\frac{1}{3}
\left(\begin{array}{ccccc}
1+ \sqrt\frac{3}{2}\,r_{1}+\frac{1}{2\sqrt{3}}\,r_{2}(1+cos^{2}\theta)& 0 & \frac{-\sqrt{3}}{2}r_{2}sin^{2}\theta  \\
 0 & 1-\frac{1}{\sqrt{3}}\,r_{2}(1+cos^{2}\theta)& 0   \\
\frac{-\sqrt{3}}{2}r_{2}sin^{2}\theta & 0 & 1- \sqrt \frac{3}{2}\,r_{1}+\frac{1}{2\sqrt{3}}\,r_{2}(1+cos^{2}\theta)
\end{array}\right)\,.
\end{equation}
\small
in $|1m\rangle$ basis.\\
The non-zero $t^{k}_{q}\,{'}s$ corresponding to this $ \rho$ are
\[t^{1}_{0} =r_{1} \,\,, \,\ t^{2}_{0} =\frac{r_{2}}{\sqrt 6}\,(1+cos^{2}\theta)\,\,,\,\,
t^{2}_{\pm 2} = \frac{-r_{2}}{2}\, sin^{2}\theta\,\,.\]

Since  $t^{1} \in \mathcal{D}^{1}_{1}$ and  $ t^{2}\in\mathcal{D}^{2}_{1,1}$ for $ 0<\theta< \frac{\pi}{2} $ and $ \frac{\pi}{2}<\theta<\pi  $, $\rho \in  \left\lbrace \mathcal{D}^{1}_{1}\,,\mathcal{D}^{2}_{1,1}\right\rbrace$. 
 
 Since $t^{1} \in \mathcal{D}^{1}_{1}$ and $ t^{2}\in\mathcal{D}^{2}_{2}$ for $\theta=0,\frac{\pi}{2},\pi$, $\rho \in \left\lbrace \mathcal{D}^{1}_{1}\,,\mathcal{D}^{2}_{2}\right\rbrace$ .

 For $r_{1}=\sqrt \frac{3}{2}$ the above density matrix is found to be pure and separable for two values of $ \theta $ only, namely $ \theta=0 $ and $ \theta=\pi $. For all other values of $ \theta $ ie; $ 0<\theta<\pi $, $ \rho $ is found to be mixed and entangled.\\
\section{CONCLUSION}
  In conclusion, we have developed a method of classifying LU equivalent classes of symmetric $N$-qubit mixed states based on the little known Multiaxial representation of the density matrix. Multiaxial representation is more general than the Majorana representation as it can be applied to pure as well as mixed states. Two states belonging to the same LU class have the same set of LU invariants and hence LU equivalent classes can also be termed as entanglement equivalent classes. Our classification is characterized by three parameters namely diversity degree, degeneracy configuration and rank. A comparative study of Majorana representation and Multiaxial representation for the $ N $-qubit GHZ state has been carried out to bring out the differences and similarities between the two representations. We have shown that in the case of GHZ states, Majorana representation is not a special case of the Multiaxial representation. Recipe for identifying $N$-qubit pure separable state is described in detail and the method is tested for some well known examples of symmetric two and three qubit pure states. We illustrate with examples, the classification of uniaxial, Biaxial and triaxial mixed states which can be produced in the laboratory. It is not clear as to why for certain configuration of the axes and certain values of $ r_{k} $, the mixed states exhibit entanglement. An indepth study of the onset of entanglement as a function of some suitable combination of LUI is needed and will be taken up in the near future.

\end{document}